\input harvmac
\newcount\figno
\figno=0
\def\fig#1#2#3{
\par\begingroup\parindent=0pt\leftskip=1cm\rightskip=1cm
\parindent=0pt
\baselineskip=11pt
\global\advance\figno by 1
\midinsert
\epsfxsize=#3
\centerline{\epsfbox{#2}}
\vskip 12pt
{\bf Fig. \the\figno:} #1\par
\endinsert\endgroup\par
}
\def\figlabel#1{\xdef#1{\the\figno}}
\def\encadremath#1{\vbox{\hrule\hbox{\vrule\kern8pt\vbox{\kern8pt
\hbox{$\displaystyle #1$}\kern8pt}
\kern8pt\vrule}\hrule}}

\overfullrule=0pt

\Title{TIFR-TH/00-52}
{\vbox{\centerline{Supergravity couplings to Noncommutative 
Branes,}
\centerline{Open Wilson lines and Generalized Star Products}}} 
\smallskip
\centerline{Sumit R.  Das~$^{a,b}$ 
\foot{das@theory.tifr.res.in} and Sandip P. Trivedi~$^a$
 \foot{sandip@theory.tifr.res.in}}
\smallskip
\centerline{$~^a${\it Tata Institute of Fundamental Research,}}
\centerline{\it Homi Bhabha Road, Bombay 400 005, INDIA.}
\smallskip
\centerline{$~^b${\it Harish-chandra Research Institute \foot{formerly
Mehta Research Institute of Mathematics and Mathematical Physics},}}
\centerline{\it Jhusi, Allahabad 211 019, INDIA.}

\bigskip

\medskip

\noindent

Noncommutative gauge theories can be constructed from ordinary
$U(\infty)$ gauge theories in lower dimensions. Using this
construction we identify the operators on noncommutative
D-branes which couple to linearized supergravity backgrounds, from a
knowledge of such couplings to lower dimensional D-branes with no $B$
field. These operators belong to a class of gauge invariant
observables involving open Wilson lines.  Assuming a DBI form of the
coupling we show, to second order in the gauge potential but to all
orders of the noncommutativity parameter, that our proposal agrees
with the operator obtained in terms of ordinary gauge fields by
considering brane actions in backgrounds and then using the
Seiberg-Witten map to rewrite this in terms of noncommutative gauge
fields.  Our result clarify why a certain {\it commutative} but {\it
non-associative} ``generalized star product'' appears both in the expansion
of the open Wilson line, as well as in string amplitude computations of
open string - closed string couplings. We outline how our procedure
can be used to obtain operators in the noncommutative theory which are
holographically dual to supergravity modes.

\Date{November 2000}

\def\cO{{\cal O}}
\def\cT{{\cal T}}
\def\tcO{\tilde{\cal O}}
\def\bA{{\bf A}}
\def\bC{{\bf C}}
\def\bX{{\bf X}}

\def\bx{{\bf x}}
\def\bp{{\bf p}}
\def\bphi{{\bf \phi}}

\def\bF{{\bf F}}
\def\bI{{\bf I}}
\def\bO{{\bf O}}

\def\sp{\star^\prime}
\def\tS{{\tilde{S}}}

\newsec{Introduction and summary}

In noncommutative gauge theory, space and color space are intertwined.
As a result there are no local position space gauge invariant observables.
However since the theories typically have translation invariance there
are such operators with definite momentum. 
These are open Wilson lines,
constructed by Ishibashi, Iso, Kawai and Kitazawa (IIKK) 
\ref\iikk{N. Ishibashi,
S. Iso, H. Kawai and Y. Kitazawa, Nucl. Phys. B573 (2000) 573,
hep-th/9910004.}. 
Consider a noncommutative Yang-Mills (NCYM) theory in 
$d+1=p+2n+1$ space-time dimensions with the
noncommutativity matrix $\theta^{AB}$ given by
\eqn\one{\eqalign{&\theta^{AB} = \theta^{ij}~~~~~~(A,B)=(i,j)~~~~
i,j = 1 \cdots 2n\cr
&\theta^{AB} = 0~~~~~~~~~~{\rm otherwise}}}
In the following we will use $i,j = 1,\cdots 2n$ to label noncommutative
directions, which are taken to be space-like,
$\mu,\nu = 1,\cdots p$ label the spatial commutative directions 
and
$A,B = 1,\cdots p+2n+1$ to label all directions collectively.
An open Wilson line $W(k,C)$ along some open contour $C$ given by 
$y^A (\lambda)$ 
with momenta $k_A$ is defined in the star product language as
\eqn\two{W (k,C) = \int d^{d+1}x~ {\rm tr}~[P_\star {\rm exp}[i\int_C d\lambda
 {d y^A(\lambda) \over d \lambda} A_A (x + y(\lambda))]~\star~e^{ik_B x^B}]}
The trace in \two\ is over the nonabelian gauge group. $\lambda$ is a parameter that increases along the path. In our conventions the path ordering is defined so that points at later 
values of $\lambda$ occur successively to  the left. Note also that all products
in \two, including those in the path ordered exponential,  are star products.
The open Wilson line \two\  is gauge invariant if the end points of the contour are separated
by an amount $\Delta x^A$ where
\eqn\three{\Delta x^A = k_B \theta^{BA}. }
Clearly the separation is nonzero only along the noncommutative directions.
When $\theta = 0$ this is just the fourier transform of
an ordinary Wilson loop with a marked point. For $\theta \neq 0$ one 
can perform a fourier
transform along the commutative directions to obtain an operator which has
a definite marked point in the commutative directions and a definite momentum
along the noncommutative directions. Various aspects of open Wilson
lines have been discussed in 
\ref\ambjorn{J. Ambjorn, Y. Makeenko, J. Nishimura
and R. Szabo, JHEP 11 (1999) 029, hep-th/9911041; J. Ambjorn,
Y. Makeenko, J. Nishimura and R. Szabo, Phys. Lett B 480 (2000) 399,
hep-th/0002158; J. Ambjorn, Y. Makeenko, J. Nishimura and R. Szabo,
JHEP 05 (2000) 023, hep-th/0004147.}.
In \ref\drey{S.R. Das and S.J. Rey,
hep-th/0008042.} it was argued that these operators (with a modification to 
include scalar fields)
form a complete set (in fact an
overcomplete set) of operators of the theory made from gauge fields and
scalars. They can be interpreted as macroscopic fundamental
strings \ref\reyunge{S.J. Rey and R. von Unge, hep-th/0007089.}.

In \ref\ghi{D. Gross, A. Hashimoto and N. Itzhaki, hep-th/0008075.}
gauge invariant operators were written down which reduce to local
gauge invariant operators in the commutative limit.
These are defined as 
\eqn\four{\tcO (k) = \int d^{d+1}x
~~{\rm tr}~~ \cO (x+ {k \cdot \theta}) \star P_\star~ {\rm exp}
[i\int_0^1 d\lambda~
k_A\theta^{AB} A_B (x + {k \cdot \theta~\lambda})]
 \star e^{ik\cdot x}.} 
Note, the contour is now a {\it straight} path transverse to the
momentum  along the direction
\eqn\five{ \eta^A  = k_B\theta^{BA}.} 
$\cO (x + {k \cdot \theta})$ is a local operator constructed from the
fields which is inserted at the endpoint of the path, and
\eqn\defdot{(k \cdot \theta)^A \equiv k_B \theta^{BA}.}

The path ordered Wilson loop factor above will be referred to as the
"tail" of the operator. It clearly extends in the noncommutative
direction.  This represents a deformation of the fourier transforms of
local gauge invariant operators to the noncommutative
theory. Correlation functions of these operators have been calculated
in \ghi\ yielding interesting results - in particular a universal
behavior at high momenta.  It is also possible to construct other
operators, e.g.  open wilson lines with self intersections and open
wilson lines ending with a closed wilson loop \ref\dw{A. Dhar and
S.R. Wadia, hep-th/0008144.}, and the operators \four\ are in fact
special cases of the latter.

Recently, a smeared version of the operator \four\ has been introduced
in \ref\liutwo{H. Liu, hep-th/0011125}. This relates to the situation
where the operator $\cO(x)$ is itself a product of operators
\eqn\wtwo{\cO(x) = \prod_{\alpha=1}^n \cO_\alpha (x)}
Then the operators $\cO_\alpha$ can be smeared over the Wilson tail 
\eqn\wone{{\hat \cO}(k)
= \int d^{d+1}x \int \prod_{\alpha=1}^n d\tau_\alpha~
~P_*~{\rm tr}~
[ \prod_{\alpha=1}^n 
O_\alpha(x^i + \theta^{ji}k_j\tau_\alpha) W_t(k,A,\phi,x)]  
\star e^{ik_ix^i}}
where $W_t$ denotes the Wilson tail
\eqn\wthree{W_t (k,A,\phi,x) 
= {\rm exp} [i\int_0^1 d\lambda~
k_A\theta^{AB} A_B (x + {k \cdot \theta~\lambda})]}

For normal $p$-branes without $B$ fields on them, coupling to a
linearized supergravity background yields a natural set of gauge
invariant operators of the worldvolume $(p+1)$-dimensional gauge
theory. The same should be true with $B$ fields and should therefore
naturally lead to a set of gauge invariant operators of the
noncommutative gauge theory, as emphasized in
\ref\dg{S.R. Das and B. Ghosh, JHEP 06 (2000) 043,
hep-th/0005007.}. Such couplings are useful in various
contexts, e.g. absorption or Hawking radiation \foot{For
absorption/radiation by black holes the coupling is sometimes to
effective theories rather than fundamental brane theories}
\ref\bhole{C. Callan and J. Maldacena, Nucl. Phys. B475 (1996) 645,
hep-th/9602043; A. Dhar, G. Mandal and S. wadia, Phys. Lett.
388B (1996) 51, hep-th/9605234;
S.R. Das and S. Mathur, Nucl. Phys. B478 (1996) 561,
hep-th/9606185; I. Klebanov, Nucl. Phys. B496 (1997) 231, 
hep-th/9702076; S. Gubser, I. Klebanov and A. Tseytlin,
Nucl. Phys. B499 (1997) 217; I. Klebanov, W. Taylor and M. van
Raamsdonk, Nucl.Phys. B560 (1999) 207, hep-th/9905174.} or discovery
of expanded brane configurations in the presence of backgrounds
\ref\bexp{R. Myers, JHEP 9912 (1999) 022, hep-th/9910053.}. A
reasonably exhaustive set of such couplings have been obtained
by matrix theory techniques in \ref\watimark{W. Taylor and
M. van Raamsdonk, Nucl. Phys. B558 (1999) 63, hep-th/9904095;
W. Taylor and M. van Raamsdonk, Nucl. Phys. B573 (2000) 703,
hep-th/9910052} and from T-duality consistencies in \bexp.

While it is obvious that these gauge invariant operators couple to
general closed string modes, so far it has not been possible to
determine in a precise fashion which operator couples to which
supergravity mode.  In this paper we propose a way to do this. 
We use the
construction of noncommutative gauge theories from ordinary
$U(\infty)$ gauge theories in lower dimensions or matrix models
\ref\kawai{H. Aoki, N. Ishibashi, S. Iso, H. Kawai, Y. Kitazawa and
T. Tada, Nucl.Phys.B565 (2000) 176, 
hep-th/9908141} \iikk\ \ambjorn\
\ref\ishibashi{N. Ishibashi, hep-th/9909176; 
N. Ishibashi, Nucl.Phys. B539 (1999)107,
hep-th/9804163}
\ref\cornalba{L. Cornalba and R. Schiappa, hep-th/9907211;
L. Cornalba, hep-th/9909081.}
\ref\seiberg{N. Seiberg, hep-th/0008013.}, which was used to write
down these operators in the first place \iikk.
We propose that once we know the linearized couplings of a set of
ordinary $Dp$ branes to supergravity backgrounds, we can use the above
construction to find the couplings of these backgrounds to
noncommutative $D(p+2n)$ branes with noncommutativity in $2n$ of the
directions.  These operators turn out to be exactly of the type
\wone\ constructed in \liutwo.

There is another way one could obtain
the couplings to noncommutative branes. One can, by direct
calculation, obtain these operators by considering the coupling of a single
closed string with several open strings in the presence of a nonzero
$B$ field and express them in terms of an ``ordinary'' gauge field
$f_{\mu\nu}$.
For example for a single noncommutative brane one may take the coupling
given by the DBI-WZ action in an arbitrary background written in terms
of ordinary gauge fields.
On the other hand, the gauge field $f_{\mu\nu}$ is related to the
noncommutative gauge field $F_{\mu\nu}$ by the Seiberg-Witten map
\ref\switten{N. Seiberg and E. Witten, JHEP 9909 (1999) 032,
hep-th/9908142.}. Using this map one can in principle obtain the
operators in terms of the noncommutative gauge fields $F_{\mu\nu}$.
One would, of course, get an infinite series and any finite truncation
would not be gauge invariant under noncommutative gauge
transformations. Furthermore the Seiberg-Witten map is not known to
all orders. Nevertheless one may carry out this procedure in an
expansion in powers of the noncommutative gauge field $A_\mu$.
This has been carried out for the DBI action in \ref\gar{M. Garousi,
Nucl. Phys. B579 (2000)  209, hep-th/9909214} where it has been
argued that the answer correctly reproduces the simplest amplitudes
involving open and closed strings obtained in \ref\hyun{S.
Hyun, Y. Kiem, S. Lee and C.Y. Lee, hep-th/9909059} and \gar.

In our proposal the operators are obtained directly in terms of the
noncommutative gauge fields and are gauge invariant by construction.
However, the answer, when expanded in powers of $A_\mu$, should agree
with the answer obtained via ordinary gauge fields and the
Seiberg-Witten map.  As a concrete check of our proposal we carry out
this comparison explicitly for the dilaton coupling to noncommutative
branes.  A nonabelian version of the Dirac-Born-Infeld
action coupled to backgrounds was discussed in \bexp\ \watimark\ \ref\mgaro{M. Garousi
and R. Myers, hep-th/0010122.}.  We assume this form  for the lower
dimensional brane used to construct the higher dimensional
noncommutative brane. We show that the resulting operator is identical,
to second order in the noncommutative gauge potential,
to the one obtained from the DBI action written in terms of ordinary
gauge fields and transformed by the Seiberg-Witten map.
For simplicity, we do the calculation where we have a single
{\it euclidean} noncommutative $(2n-1)$ brane which is obtained from
the action of an infinite number of D-instantons (in DBI
form). However the calculation may be easily generalized to lorentzian
branes (with magnetic type $B$ fields on them). Extension of our
results to arbitrary number of noncommutative branes requires a
solution to the Seiberg-Witten map for nonabelian gauge fields.

The solution to the Seiberg-Witten map yields an interesting structure
: the result appears in terms of a ``generalized star product'' which
are {\it commutative} but {\it non-associative} \gar and a triple
product \ref\mw{T. Mehen and M. Wise, hep-th/0010204.}. These
generalized products therefore appear in the open-closed string
couplings as well.  The same generalized products appear in one loop
effective actions of NCYM theories \ref\oneloopa{H. Liu and
J. Micheslon, hep-th/0008205}
\ref\oneloopb{D. Zanon, hep-th/0009196;
A. Santambrogio and D. Zanon, hep-th/0010275} and in the
study of anomalies \ref\AS{F. Ardalan and N. Sadoghi,
hep-th/0009233.}. Recently it has been shown \mw,\liutwo\
that these generalized
products {\it also} appear in the expansion of the open Wilson lines
considered in \iikk\ -
\dw. Our result therefore provides an explanation as to why the same
structure appears in open-closed interactions as well in the gauge
invariant open Wilson lines \foot{In \gar\ it was proposed that to
obtain the coupling of a mode to a noncommutative brane (in the DBI
approximation) one has to first write down the usual coupling,
replace ordinary products by generalized star products and then
use the Seiberg-Witten map. We have not been able to see why this
prescription is correct.}.

Gauge invariant operators also appear in the context of holography.
The states created by such operators would have a  dual description as
normal modes in the dual supergravity background. It turns out that
the asymptotic geometry for the $p+2n+1$ dimensional non-commutative theory is 
{\it identical} to that for the $p+1$ dimensional ordinary theory at a particular point in the Coulomb branch where  the $p$-branes are spread out uniformly along the $2n$ directions.
This is in fact the
dual manifestation of the relationship between commutative and
noncommutative Yang-Mills theories discussed above 
\ref\roylu{J.X. Lu and S. Roy, Nucl.Phys. B579 (2000) 229,
hep-th/9912165; R. Cai and N. Ohta, JHEP 0003 (2000) 009,
hep-th/0001213.}.This connection may be possibly used to 
tackle the problem of mode mixing in such supergravity backgrounds.
We do not have definitive results about this at present.

\newsec{Noncommutative Yang-Mills from lower dimensional ordinary Yang-Mills}

In this section we review how noncommutative Yang-Mills theories are
obtained from lower dimensional ordinary $U(\infty)$ Yang Mills
theories.  This is how how space-time emerges in Eguchi-Kawai models
\ref\ek{T. Eguchi and H. Kawai, Phys. Rev. Lett. 48 (1982) 1063;
G. Parisi, Phys. Lett. 112B (1982) 463; G. Bhanot, U. Heller and
H. Neuberger, Phys. Lett 113B (1982) 47; D. Gross and Y. Kitazawa,
Nucl. Phys. B 206 (1982) 440; S.R. Das and S.R. Wadia,
Phys. Lett. 117B (1982) 228; S.R. Das, Rev. Mod. Phys. 59 (1987)
235.}, particularly in its ``twisted'' version,
\ref\tek{A. Gonzalez-Arroyo and M. Okawa, Phys. Rev. D27 (1983) 2387;
T. Eguchi and R. Nakayama, Phys. Lett. 122B (1983) 59;
A. Gonzalez-Arroyo and C. Korthals-Altes, Phys. Lett. 131B (1983)
396.} and how branes arise in matrix models \ref\nicolai{B. deWit,
J. Hoppe and H. Nicolai, Nucl. Phys B305 (1989) 135.}.  In modern
Matrix theory, both of the BFSS \ref\matrix{T. Banks, W. Fischler,
S. Shenker and L. Susskind, Phys. Rev. D55 (1997) 5112,
hep-th/9610043.} as well as the IKKT type \ref\ikkt{N. Ishibashi,
H. Kawai, Y. Kitazawa and A. Tsuchiya, Nucl. Phys. B498 (1997) 467,
hep-th/9612115.} branes arise in a similar way
\ref\matrixbranes{O.J. Ganor, S. Ramgoolam and W. Taylor,
Nucl. Phys. B492 (1997) 191, hep-th/9611202; T. Banks, N. Seiberg and
S. Shenker, Nucl. Phys. B490 (1997) 91, hep-th/9612157.}. This has led
to the discovery of noncommutativity in string theory
\ref\cds{A. Connes, M. Douglas and A. Schwarz, JHEP 9802 (1998) 003,
hep-th/9711162; M. Douglas and C. Hull, JHEP 9802 (1998) 008,
hep-th/9711165.}  and has been useful in providing valuable insights
\ref\schomerus{ V. Schomerus, JHEP 9906 (1999) 030}\switten.
Several useful aspects of this connection are contained in
\iikk\ \ambjorn\ \ishibashi\ \ref\li{M. Li, Nucl. Phys. B499 (1997) 149, hep-th/9612222.}
\ref\bars{I. Bars and D. Minic, hep-th/9910091} \ref\luis{L. Alvarez-Gaume and
S.R. Wadia, hep-th/0006219; A.H. Fatollahi, hep-th/0007023.}.
We will use the framework of \kawai\ and \seiberg.

Consider a $U(\infty)$ ordinary gauge theory in $(p+1)$ dimensions
with the usual gauge fields $\bA_\mu(\xi)~~,\mu=1,\cdots p+1$ and $(9-p)$
scalar fields $\bX^I(\xi),~I=1,\cdots (9-p)$ in the adjoint representation,
together with their fermionic partners. In this paper we will restrict ourselves to only bosonic components of operators, consequently, fermions will not enter the subsequent discussion.
The bosonic part of the action is
\eqn\six{S = {\rm Tr}\int d^{p+1}\xi [\bF_{\mu\nu}\bF^{\mu\nu} + 
D_\mu \bX^I D^\mu \bX^J g_{IJ}
+ [\bX^I,\bX^J][\bX^K,\bX^L]g_{IK}g_{JL}]}
where $g_{IJ}$ are constants and the other notations are standard. Boldface
has been used to denote $\infty \times \infty$
matrices. 

The action has a nontrivial classical solution
\eqn\seven{\eqalign{& \bX^i (\xi) = \bx^i ~~~~~~~~~~i=1,\cdots 2n\cr
&\bX^a=0~~~~~~~~~~a=2n+1 \cdots 9-p\cr
&\bA_\mu = 0}}
where the constant (in $\xi$) matrices $\bx^i$ satisfy
\eqn\eight{[\bx^i,\bx^j] = i\theta^{ij} \bI}
The antisymmetric matrix $\theta^{ij}$ has rank $p$ and $\bI$
stands for the unit $\infty \times \infty$ matrix. The inverse of the
matrix $\theta^{ij}$ will be denoted by $B_{ij}$

The idea is then to expand the various fields as follows.
\eqn\nine{\eqalign{&\bC_i = B_{ij}\bX^j = \bp_i + \bA_i\cr
&\bX^a = {\bphi}^a \cr
&\bA_\mu= \bA_\mu}}
where
\eqn\ten{\bp_i = B_{ij}\bx^j}

We will expand any matrix $\bO(\xi)$ as follows
\eqn\twelve{\bO(\xi) = \int d^{2n}k~{\rm exp}[i\theta^{ij}k_i\bp_j]~
O(k,\xi)}
where $O(k,\xi)$ are ordinary functions. Regarding these $O(k,\xi)$
as fourier components of a function $O(x,\xi)$, where $x^i$ are the
coordinates of a $2n$ dimensional space we then get the following
map between matrices and functions.
\eqn\eleven{\eqalign{&\bO (\xi) \rightarrow O(x,\xi)\cr
&[\bp_i,\bO (\xi)] = i\partial_i O(x,\xi)\cr 
&{\rm Tr}\bO (\xi) = {1
\over (2\pi)^{n}}[{\rm Pf~B}]\int d^{2n}x~~O(x,\xi)}} 
The product of two
matrices $\bO_1 (\xi)$ and $\bO_2 (\xi)$ is then mapped to a star
product
\eqn\thirteen{ \bO_1(\xi) \bO_2(\xi) 
\rightarrow O_1(x,\xi)*O_1(x,\xi)}
where
\eqn\fourteen{O_1(x,\xi)*O_2(x,\xi)
= {\rm exp}~[{\theta^{ij}\over 2i}{\partial^2 \over \partial s^i
\partial t^j}]~O_1(x+s,\xi)O_2(x+t,\xi)~|_{s=t=0}}
A quick way to see this is to consider the operators 
\eqn\ffone{\bO (k) = {\rm exp}~[i\theta^{ij}k_i\bp_j]}
which form a complete basis. Then the commutation relations of
$\bx^i$ and hence $\bp_i$ show
\eqn\fftwo{\bO(k) \bO(k') = e^{-{i\over 2} \theta^{ij}k_ik'_j}~
\bO(k+k')}

With these rules, one can easily verify
\eqn\fifteen{\eqalign{& \bF_{\mu\nu} \rightarrow
\partial_\mu A_\nu - \partial_\nu A_\mu 
-i A_\mu *A_\nu + i A_\nu *A_\mu \equiv F_{\mu\nu}\cr
& D_\mu \bX^i \rightarrow \theta^{ij}(\partial_\mu A_j 
-\partial_j A_\mu  -i A_\mu * A_j
+ i A_j * A_\mu) \equiv \theta^{ij}F_{\mu j}\cr
&D_\mu \bX^a \rightarrow \partial_\mu \phi^a  - iA_\mu *
\phi^a  + i \phi^a *A_\mu \equiv D_\mu \phi^a\cr
&[\bX^i,\bX^j] \rightarrow i\theta^{ik}\theta^{jl}(F_{kl}-B_{kl})\cr
&[\bX^i,\bX^a] \rightarrow i\theta^{ij}(\partial_j \phi^a - iA_j \star \phi^a
+ i \phi^a \star A_j) \equiv i\theta^{ij}D_j\phi^a}}
where we have defined
\eqn\sixteen{F_{ij} = \partial_i A_j-\partial_jA_i - 
iA_i\star A_j + iA_j\star A_i}
In the above equations the quantities appearing in the right hand side
are ordinary functions of $(x,\xi)$.

The action \six\ becomes the
action of $U(1)$ noncommutative gauge theory in the $p+2n+1$ dimensions
spanned by $x,\xi$. The noncommutativity is entirely in the $2n$
directions. In addition to the gauge fields we also have $(9-p-2n)$
``adjoint'' scalars $\phi^a$. 
The gauge field appears in the combination
\eqn\seventeen{ F_{AB} - B_{AB}}
where $B_{AB}$ is an antisymmetric matrix whose $(ij)$ components are
$B_{ij}$ and the rest zero. This corresponds to a specific choice of
``description'' in the NCYM theory \seiberg. Furthermore the upper and
lower indices of various quantities some contracted with the ``open
string metric'' whose components in the nocommutative directions are
\eqn\seventeena{G^{ij} = - \theta^{ik}g_{kl}\theta^{lj}}
The componments of the open string metric in the commutative directions
are the same as the original metric $g_{ab}$. Finally the coupling
constant which appears in front is the open string coupling $G_s$ which
is related to the closed string coupling $g_s$ by
\eqn\seventeenb{ G_s = g_s ({{\rm det}(G-B)\over{\rm det}(g+B)})^{1\over 2}
= g_s ({{\rm det}~B\over{\rm det}~g})^{1\over 2}} 
It may be also easily verified that
\eqn\dilseven{ {1\over G-B} = -\theta + {1\over g+B}}
(Recall that $\theta^{-1} = B$ as matrices.)

It is straightforward to extend the above construction to obtain a
nonabelian noncommutative theory. The classical solution which one
starts with is now
\eqn\eighteen{\bX^i (\xi) = \bx^i \otimes I_M}
where $I_M$ denotes the unit $M \times M$ matrix. Now the various
$\infty \times \infty$ matrices map on to $M \times M$ matrices which
are functions of $x$, in addition to $\xi$. 
With this understanding the formulae above can
be almost trivially extended. The star product would now include
matrix multiplication and the map for the trace becomes
\eqn\nineteen{{\rm Tr}\bO (\xi) = {1
\over (2\pi)^{n}}[{\rm Pf~B}]\int d^{2n}x~~{\rm tr}~O(x,\xi)}
where ${\rm tr}$ now denotes trace over $M \times M$ matrices. Instead
of obtaining a $U(1)$ noncommutative theory one now obtains a
$U(m)$ noncommutative theory.

Finally the expression for the open Wilson line \two\ is easily seen
to be \iikk\
\eqn\twenty{\eqalign{&W(C,k) = \int d^{p+1}\xi
~{\rm Lim}_{M \rightarrow \infty}~~
{\rm Tr}~[\prod_{n=1}^M~U_j]~e^{ik_\mu \xi^\mu}\cr
& U_j = {\rm exp}~[i {\vec \bC} \cdot ({\vec \Delta d})_n]}}
where $\Delta d_n$ denotes the $n$-th infinitesimal line element along
the contour $C$. The momentum components $k_\mu$ along the
commutative directions appear explicitly
in \twenty, while the components along the noncommutative directions
$k^i$ are given by
\eqn\twoone{k_i = B_{ji}d^j}
where $d^j$ are the components of the vector
\eqn\twotwo{{\vec d} = \sum_{n=1}^M {\vec{\Delta d}}}

Operators with straight Wilson line tails given by \four\ are similarly
represented by 
\eqn\twothree{\cO(k) = \int d^{p+1}\xi~e^{ik_\mu \xi^\mu}~{\rm Tr}~
[e^{ik_i\bX^i}~ \cO(\bX,\bA,\xi)]}

\newsec{Branes in supergravity backgrounds}

Consider a large number of coincident $p$ branes with no $B$ field in
the presence of a weak supergravity background. Let us denote a 
supergravity mode in momentum space by $\Phi (k_I,k_\mu)$ where $k_\mu$ denotes the
momentum along the brane and $k_I$ denotes the momentum transverse to the
brane. Let $X^I$ denote the transverse coordinate and $\bA_\mu$ the gauge
field on the brane. Then in the brane theory, the transverse coordinates
are represented by scalar fields $\bX^I(\xi)$. Suppose 
a linearized coupling of the
mode to the set of branes is given by
\eqn\twofour{\Phi (k_\mu,k_I) \int d^{p+1}\xi~e^{ik_\mu\xi^\mu}~{\rm Tr}~
[e^{ik_I\bX^I}~\cO_\phi(\bX,\bA,\xi)]}
Such operators can be derived by various methods, for example by using
T-duality on Matrix Theory results \watimark.  Note, this is a
coupling to an operator quite similar to \twothree.  The only
difference is that among the $\bX^I$'s some of them, $\bX^a$ are
expanded around the trivial solution, while the $\bX^i$ are expanded
around the nontrivial solution $\bx^i$. A straightforward extension of
the manipulations performed in \iikk\ allows us to rewrite this in the
language of functions and star products. This leads to the
generalization of the Wilson line given in \drey. The final expression
for the coupling of the {\it same} supergravity mode to a set of $M$
noncommutative $(p+2n+1)$ branes
\eqn\twofive{\eqalign{S_{int} & = \Phi(k)\int {d^{p+1}\xi~d^{2n}x 
\over (2\pi)^n}~
({\rm Pf}B)~e^{ik_\mu \xi^\mu}~{\rm tr}~
[O_\phi(x + {k \cdot \theta},\xi) 
\star P_\star(W_t(k,A,\phi))  \star e^{ik_ix^i}]\cr
& W_t(k,A,\phi) = 
{\rm exp}[i\int_0^1 d\lambda ~k_i\theta^{ij} A_j (x + \eta(\lambda))
+ i
\int_0^1 d\lambda ~k_a \phi^a(x + \eta (\lambda))]}}
where $\eta^i(\lambda) = \theta^{ji}k_j\lambda$. 
The operator $O_\phi (x,\xi)$
is obtained from ${\bf{\cO}}_\phi$ by the mapping discussed
in the previous section. In the rest of the paper we will set $k_a=0$
for simplicity, so that the supergravity mode has no momentum in
directions transverse to the resulting noncommutative brane. A
nonzero $k_a$ can be restored easily using the above formula.

The coupling to the branes is, however, not quite given by \twofour
\ \foot{Some of the arguments in this section arose out a 
conversation with M. van Raamsdonk.}.
It was found in \watimark\ that the trace appearing in \twofour\ is
in fact a {\it symmetrized} trace defined as follows 
\foot{A similar symmetrized trace
appears in Tsyetlin's prescription for the nonabelian DBI action
\ref\tseytlin{A. Tseytlin, Nucl. Phys. B501 (1997) 41, hep-th/9701125.}}.
The operator $\cO_\phi (\bX,\bA,\xi)$ is in general a composite operator
made out of field strengths,
$\bF_{\mu\nu}$, the covariant derivatives of the scalar fields $D_\mu\bX^I$
and $[X^I,X^J]$. That is
\eqn\kofour{\cO_\phi (\bX,\bA,\xi)
= \prod_{\alpha=1}^n \cO_\alpha (\bX,\bA,\xi)}
where each of the $\cO_\alpha$ denotes a $\bF_{\mu\nu}$, $D_\mu\bX^I$
or a $[X^I,X^J]$.
Then imagine expanding the exponential in $e^{i k \cdot \bX}$ in
\twofour. For some given term in the exponential we thus have a 
product of a number of $\bX$'s and $\cO_\alpha$'s. Finally we 
symmetrize these various factors of $\bX$'s and $\cO_\alpha$'s 
and average. The resulting symmetrized trace will be denoted by
the symbol ``STr'' below. The coupling is then of the form
\eqn\twofoura{\Phi (k_\mu,k_I) \int d^{p+1}\
\xi~e^{ik_\mu\xi^\mu}~{\rm STr}~
[e^{ik_I\bX^I}~
\prod_{\alpha=1}^n \cO_\alpha (\bX,\bA,\xi)]}
Following the same steps as above, it is straightforward to write down
the corresponding operator in the star product language.  Then the
effect of symmetrized trace is to place these various operators
$\cO_\alpha$ 
along the path $C$ defining the Wilson line and performing a
path ordering. The final result is an operator of the form \wone,
\eqn\kofive{S_{int}  = \Phi(k)\int {d^{p+1}\xi~d^{2n}x 
\over (2\pi)^n}~
({\rm Pf}B)~\int \prod_{\alpha=1}^n d\tau_\alpha~ e^{ik_\mu
\xi^\mu}~{\rm tr}~P_\star [ \prod_{\alpha=1}^n O_\alpha(x^i +
\theta^{ji}k_j\tau_\alpha) W_t]
\star e^{ik_ix^i}}
where $W_t$ has been defined above. This is precisely an operator
of the form \wone.

These couplings are at the linearized level and the backgrounds
produced by the branes are ignored, which is the situation at weak
string coupling.  This means one can couple the $p$ brane to any
on-shell supergravity fluctuation about flat space.  We see that once
this coupling is known the coupling to the $p+2n$ brane is determined
uniquely.

\newsec{Generalized star products}

The generalized star product is defined by
\eqn\plone{f(x) \star^\prime g(x) = 
{\sin({\partial_1 \wedge \partial_2 \over 2})\over
{\partial_1 \wedge \partial_2 \over 2}}~f(x_1) g(x_2) |_{x_1=x_2=x}}
and the triple product is defined by
\eqn\pltwo{[f(x)g(x)h(x)]_{*3} = 
[ {\sin({\partial_2\wedge \partial_3 \over 2})
\sin({\partial_1 \wedge(\partial_2+\partial_3) \over 2})
\over {(\partial_1+\partial_2)\wedge \partial_3 \over 2}~
{\partial_1\wedge ( \partial_2 + \partial_3) \over 2}}
+ (1 \leftrightarrow 2) ]~f(x_1)g(x_2)h(x_3) |_{x_1=x_2=x_3=x}}
where
\eqn\plthree{
\partial_1 \wedge \partial_2 = \theta^{i j} {\partial \over \partial x^i_1}
 {\partial \over \partial x^j_2} \, }
This $\star^\prime$ product is symmetric in $f$ and $g$ and $\star_3$
is invariant under all permutations of $f,g$ and $h$ \mw.  However
\eqn\plfour{ (f(x) \sp g(x))\sp h(x) \neq f(x) \sp (g(x) \sp h(x))}
so that it is {\it commutative but nonassociative}. Nevertheless it may be
verified that
\eqn\plfive{\int dx (f(x) \sp g(x))\sp h(x)
= \int dx f(x) \sp (g(x) \sp h(x))}
so that if the generalized star product appears in an action it does
make sense. If one of the three functions in the product $[fgh]_{*3}$
is a constant this reduces to a $\sp$ product
\eqn\plfivea{[A~g(x)h(x)]_{*3} = A g(x) \sp h(x)~~~~~~A = {\rm constant}}

Another property of the $\sp$ product will be useful in the
following \mw\
\eqn\pqone{\theta^{ij}\partial_i f \sp \partial_j g = -i(f*g - g*f)}

As shown in \mw, these  generalized star products and triple products
appear in the expansion of the gauge invariant Wilson
line in powers of $A$. This may be seen by directly expanding the
expression \four\ or equivalently \twothree. 
The following identity is responsible for the appearance of the
$\star^\prime$ product :
\eqn\plsix{\int_0^1 d\sigma  \bO(k) \bO(k') e^{i(k \wedge k')\sigma}
= {\sin ({k \wedge k' \over 2})\over {k \wedge k' \over 2}}~
\bO(k + k')}
where the operators $\bO (k) $ have been defined in \ffone. This
leads to the identity
\eqn\plseven{\int d\sigma e^{i(k \wedge k')\sigma}~e^{ikx} \star e^{ik'x}
= e^{ikx} \sp e^{ik'x}}
The expansion of $\tcO (k)$ (defined in \four) 
to second order in $A$ is (for $k_a = 0$)\mw,
\eqn\pleight{\cO(k) =
\int d^{p+1}\xi {d^{2n}x \over
(2\pi)^n} e^{ik_\mu\xi^\mu}~({\rm PfB}){\rm tr}~[\cO (x,\xi)
+ \theta^{ij}\partial_j(\cO \star^\prime A_i) + 
 {{1\over 2}} \theta^{ij}
\theta^{kl}\partial_j\partial_l[\cO A_i A_k]_{*3}]\star
e^{ik_ix^i}} 
One would expect that at higher orders different
structures will emerge.

The operators which are obtained from symmetrized traces as in
\twofoura\ and \kofive\ can be similarly expanded in terms
of generalized products \liutwo.

In a sense the generalized star product is not a fundamentally
different structure : it appears because of the integration over the
parameter $\sigma$ in \plsix. If we retain this integral the answer is
always written in terms of the conventional star product. However it
is more natural to perform the $\sigma$ integral to make the operator
look local in position space : in that case the $\star^\prime$ product
appears. It should be emphasized that whatever the notation, the
operator is actually nonlocal in position space - since translations
are equivalent to specific noncommutative gauge transformations
\drey,\ghi\ and the operator has to be gauge invariant.

These generalized star products also appear in the explicit solution
of the Seiberg Witten map which relates an ordinary gauge field
$f_{ij}$ to the noncommutative gauge field $F_{ij}$ and the 
noncommutative gauge potential $A_i$. In a $U(1)$ theory the map
is given, upto two powers of $A_i$, by \gar\
\eqn\plnine{f_{ab} = F_{ab} + \theta^{kl}(A_k \star^\prime 
\partial_l F_{ab} - F_{ak} \star^\prime F_{bl}) + O(A^3)} 
Again to higher orders the triple product appears \mw. As explained
in the introduction this leads to the apperance of these products
in the closed string - open string interactions when expressed 
in a power series in $A_i$.

Our proposal for operators which couple to supergravity modes then
provides a natural explanation why these same products also appear in
the direct string amplitude calculations of \hyun\ and \gar. This is
simply because these operators are precisely appropriate momentum
space operators with straight Wilson tails.

Our proposal also explains why the generalized star products appeared
in one loop effective action calculations \oneloopa,
\oneloopb\ in the first
place. For usual gauge theories, this one loop effective action for
the massless fields obtained in the symmetry breaking $U(N_1+N_2)
\rightarrow U(N_1) \times U(N_2)$ can be alternatively viewed as the
potential between a set of $N_1$ branes and another set of $N_2$
branes separated by a distance \ref\branewave{M. Douglas, D. Kabat,
P. Pouliot and S. Shenker, Nucl. Phys. B485 (1997) 85, hep-th/9608024;
M. Douglas and W. Taylor, hep-th/9807225; S.R. Das, JHEP 9902 (1999)
012, hep-th/9901004; S.R. Das, JHEP 9906 (1999) 029, hep-th/9905037.}
due to exchange of supergravity modes. If the same is true for
noncommutative gauge theories, generalized products in the
supergravity couplings naturally lead to their presence in the
effective action.

\newsec{Dilaton couplings to noncommutative branes}

In this section we perform a test of our proposal. We consider 
the coupling of the
dilaton to noncommutative branes in a DBI approximation 
and show that our proposal is
consistent with the opertors which would be obtained by starting
out with ordinary gauge fields and using the Seiberg Witten map.

For simplicity we consider a {\it single} noncommutative {\it euclidean} 
$D(2n-1)$ brane (with $2n$ dimensional worldvolume) and we will
construct this from a large number $N$ of $D(-1)$ branes. Following
\bexp,\watimark and \mgaro\ we will assume that the action in the
presence of a dilaton field $D(x)$ (with all backgrounds trivial)
is given by
\eqn\dilone{S = {1\over g_s} {\rm STr}~e^{-D(\bX^I)}{\sqrt{{\rm det}
(\delta^I_J - i [\bX^I,\bX^K]g_{KJ})}}}
Here, as before, $X^I$ denote all the $d$ transverse coordinates. $g_{IJ}$
is a constant closed string metric which is taken to be diagonal as 
well \foot{the diagonal nature of the closed string metric has been
used to arrive at \dilone\ starting from the form of action given e.g.
in \bexp.} and $g_s$ is the closed string
coupling. The meaning of the symmetrized trace has been explained in
the previous section.
As in 
the previous sections we will write the background in terms of its
fourier transform so that for a given space-time momentum $k$ the
linearized coupling is 
\eqn\diltwo{S_{int} = {D(k)\over g_s} {\rm STr}~
e^{ik\cdot \bX}{\sqrt{{\rm det}
(\delta^I_J - i [\bX^I,\bX^K]g_{KJ})}}}
where $D(k)$ is the fourier transform of $D(x)$
The classical solution which leads to a noncommutative
$(2n-1)$ brane
\eqn\dilthree{\eqalign{& \bX^i = x^i ~~~~~i = 1 \cdots 2n \cr
& \bX^a = 0~~~~~a= (2n+1) + \cdots d}}
We then expand around this
classical solutions as in (2.3) and (2.4).
To simplify things further we will assume that $k_a = 0$ so that
we have dependence only on $X^i$ and also set the scalar fields to
zero, $\phi^a = 0$. It is trivial to repeat the following for 
nonzero $k_a$ and $\phi^a$.

In the following we will be interested in terms upto $O(A^2)$ in the
noncommutative gauge fields. In the language of matrices we will be
interested in terms which contain at most two matrices. For such terms
there is no distinction between the symmetrized trace and ordinary
trace. We will therefore replace STr in \diltwo\ with Tr.
Using the results of section 2, this interaction is then written
in terms of noncommutative gauge fields $F_{ij}$ 
\eqn\dilfour{S_{int} = {D(k)\over g_s}|{\sqrt{{\rm det}B}}|
\int d^{2n}x~e^{ikx}~P_*[{\rm exp}~(i\int d\eta^i A_i (x + \eta(\lambda)))]
~{\sqrt{{\rm det}
(I - \theta (F-B) \theta g)}}}
where in \dilfour\ the quantities $\theta,F,B,g$ are written as $(2n)
\times (2n)$ matrices and $I$ stands for the identity matrix, 
in a natural notation. In the following whenever these
quantities appear without indices they denote these matrices. We now use
\seventeena\ and \seventeenb\ to write this in terms of the open string
metric $G_{ij}$ and the open string coupling $G_s$ as
\eqn\dilfive{S_{int} = {D(k)\over G_s}
\int d^{2n}x~e^{ikx}~P_*[{\rm exp}~(i\int d\eta^i A_i (x + \eta(\lambda)))]
~{\sqrt{{\rm det}
(G + F - B)}}} 
Here the path used is given by \five\ and all products are star products.

In terms of the ordinary gauge fields $f_{ij}$, the closed string metric
and the closed string coupling, the interaction may be
read off from the standard Dirac-Born-Infeld action
\eqn\dilsix{\tS_{int} = {D(k)\over g_s}
\int d^{2n}x~e^{ikx}~{\sqrt{{\rm det}
(g + f +B)}}} 

The strategy is now to express \dilsix\ in terms of the noncommutative
gauge field $F_{ij}$ using the Seiberg-Witten map in a series involving
powers of the potential $A_i$ and compare the result with \dilfive\
which is also expanded in a similar fashion. 

For zero momentum operators this is the comparison done in \switten,
where it is shown that
\eqn\dilsixa{{1\over g_s}{\sqrt{{\rm det}
(g + f +B)}} = {1\over G_s}{\sqrt{{\rm det}
(G -B + F)}} + O(\partial F) + {\rm total~derivatives}}
which shows the equivalence of the two actions in the presence of 
constant backgrounds. The crucial aspect of our comparison is the presence
of these total derivative terms in \dilsixa, which cannot be ignored
if $k \neq 0$. We will find that these total derivative terms are
in precise agreement with similar terms coming from the expansion of the
Wilson tail in \dilfive, upto $O(A^2)$.

Since we are using the DBI action, the field strengths should be really
treated as constant.  In carrying out the comparison though some caution must be exercised. 
Since the Seiberg-Witten map contains gauge
potentials as well as field strengths a term containing  a derivative of
a field strength multiplied by a gauge potential without a
derivative on it,  cannot be set automatically to zero, as emphasised in \switten.

\subsec{$O(A)$ comparison}

First let us do the comparaison to $O(A_i)$. To this order the 
Seiberg-Witten map in \plnine\ simply reduces to
$f_{ij} = \partial_i A_j - \partial_j A_i + O(A^2)$.
Thus it is sufficient to expand the determinant in \dilsix\ 
to linear order in
$f$. One obtains
\eqn\dilnine{\tS^{(1)}_{int} = { D(k) \over g_s}\sqrt{{\rm det}
(g + B)}~\int d^{2n}x ~e^{ikx}~
[1 + {1\over 2} ({1\over (g+B)})^{ij}(\partial_j A_i 
- \partial_i A_j) + O(A^2)]}
Using \seventeenb\ and \dilseven\ this may be written as
\eqn\dilten{\tS^{(1)}_{int} = { D(k) \over G_s}
\sqrt{{\rm det}(G - B)}~\int d^{2n}x ~e^{ikx}~[1 +
{1\over 2} ({1\over (G-B)} + \theta )^{ij}(\partial_j A_i 
- \partial_i A_j)+ O(A^2)]}

We have to compare this with the expansion of the expression
\dilfive\ to $O(A)$. In this expression all products are star
products. To do this we can use \pleight\ with the function
$\cO$ being replaced by the quantity ${\sqrt{{\rm det}
(G + F - B)}}$. To linear order in $A$ we have
\eqn\dileleven{{\sqrt{{\rm det}
(G + F - B)}}= {\sqrt{{\rm det}(G-B)}}[1 + {1\over 2}
({1\over G-B})^{ij} (\partial_i A_j 
- \partial_j A_i) + O(A^2)]}
The various products appearing on the left hand side of
the above equation are star products. However to this order
these collapse to ordinary products since $G,B$ etc. are
constants.
Also to this order one has
\eqn\diltwelve{ \theta^{ij}\partial_j ({\sqrt{{\rm det}
(G + F - B)}} \sp A_i) = \theta^{ij}{\sqrt{{\rm det}
(G - B)}} \partial_j A_i = 
{{1\over 2}} {\sqrt{{\rm det}
(G - B)}} \theta^{ij}(\partial_j A_i - \partial_i A_j)}
Thus substituting \dileleven\ and \diltwelve\ in $S_{int}$
we have after using (4.8)
\eqn\dilthirteen{S^{(1)}_{int}=
{D(k)\over G_s}
\sqrt{{\rm det}(G - B)}~\int d^{2n}x e^{ikx}~[1 +
{1\over 2} ({1\over (G-B)} + \theta )^{ij}(\partial_j A_i 
- \partial_i A_j) + O(A^2)]}
which is exactly the same as \dilten.

Note that the term proportional to $\theta$ in \dilten\ came
because of the relation \dilseven, while the corresponding
term in \dilthirteen\ came from the ``Wilson tail'' involved
in the gauge invariant operator. To this order one is sensitive only to the linear term
of the Seiberg Witten map.  However the agreeement
of the two derivations of the interaction term is still nontrivial
and the importance of the open Wilson line is evident.

\subsec{$O(A^2)$ comparison}

To next order, several points have to be remembered. First the
Seiberg-Witten map is nontrivial. Secondly, we have to be careful
about where we can ignore star products and where we can not. 
The strategy is the same as in the previous subsection. We expand
the expression in terms of ordinary gauge fields to the required
order and reexpress the terms using Seiberg Witten map, after
using \dilseven. Finally we compare the resulting expression with
the expansion of \dilfive. 
It will be necessary to write the noncommutative gauge field strength
\sixteen\ in terms of $\sp$ products by using \pqone,
\eqn\dilfourteen{F_{ij}=\partial_iA_j-\partial_jA_i 
+\theta^{kl}(\partial_k A_i \sp \partial_l A_j)}

The details of the calculation are given in the Appendix. Here we quote
the final result. The expansion of \dilsix\ becomes
\eqn\dilfifteen{\eqalign{{\tS_{int}\over D(k)} =
{{\sqrt{{\rm det}(G - B)}} \over G_s}\int d^{2n}x~& e^{ikx}[1 +
{1\over 2} M^{ij}F_{ji} - {1\over 4}M^{ij}F_{jk}
M^{kl}F_{li}
+ {1\over 8}M^{ij}F_{ji}M^{kl}F_{lk}\cr
& + \theta^{ij}\partial_jA_i + {1\over 4}\theta^{ij}F_{ji}
M^{kl}F_{lk} + {{1\over 2}} \theta^{ij} M^{kl}
(\partial_j F_{lk}
\sp A_i) \cr
& + {{1\over 2}} \theta^{ij}\theta^{kl}(\partial_l F_{ji} \sp A_k)
+ {1\over 8} \theta^{ij} F_{ji}\theta^{kl} F_{kl} \cr
& + {1\over 2} \theta^{ij} \theta^{kl} (\partial_l A_i
\sp \partial_j A_k) + O(A^3)]}}
where we have defined
\eqn\dilsixteen{ M^{ij} = ({1\over G-B})^{ij}}
The result is {\it exact to all orders in }$~\theta$, but 
to $O(A^2)$.

On the noncommutative side, \dilfive\ may be written as
\eqn\dilseventeen{S_{int}= {D(k)\over G_s}\int d^{2n}x~e^{ikx}[
P(x) + \theta^{ij}\partial_j(P(x) \sp A_i) + 
{{1\over 2}} \theta^{ij}\theta^{kl}
\partial_j \partial_l [P(x)A_i(x)A_k(x)]_{*3}] + O(A^3)}
where
\eqn\dileighteen{P(x) = {\sqrt{{\rm det}
(G + F(x) - B)}}}
Here, in the expansion of the determinant in powers of $F$ we can
replace the star product by ordinary products. We perform this expansion
to the requisite power of $A$ and find that the result is in exact agreement
with \dilfifteen.

It should be possible to extend the discussion in this section can be
extended to any other mode, to include fluctuations of the $\phi^a$
and to Lorentzian signature.
It will be particularly interesting to study the couplings to the
RR fields : these Chern-Simons couplings for {\it constant} backgrounds
have been obtained in \ref\ms{S. Mukhi and N. Suryanarayana,
hep-th/0009101.} and one has to extend this to nonconstant backgrounds
using the proposal of this paper.

\subsec{Higher orders}

To higher order terms in $A$, the difference between symmetrized trace
and ordinary trace becomes important. Because of this, it is important
to extend the above comparison to higher orders. We have not performed
this calculation yet.

\subsec{Other Descriptions}
Let us make one comment before proceeding.
Constructing the noncommutative Yang Mills theory
from the $D(-1)$ branes  leads (in the notation of \switten\ )
to the $\Phi =-B$ description.
One conjecture for the coupling of the Dilaton in a general description is:
\eqn\dilfive{S_{int} = {D(k)\over G_s}
\int d^{2n}x~e^{ikx}~P_*[{\rm exp}~(i\int d\eta^i A_i (x + \eta(\lambda)))]
~{\sqrt{{\rm det}
(G + F +\Phi)}}}
The $\Phi=B$ description corresponds to the case with ordinary gauge field $f_{ij}$,
 closed string metric and closed string coupling  and to the dilaton coupling  \dilsix. 

Above we showed that upto $O(A^2)$ the coupling in the 
$\Phi=-B$ description agreed with that in the $\Phi=B$ case. 
It is interesting to note that the calculations can be repeated for other descriptions,
i.e. other values of $\Phi$ in a very straightforward fashion. 
One finds that  the 
ansatz \dilfive\  in other descriptions  agrees with the coupling \dilsix\
  as well, upto $O(A^2)$. 
This follows in a straightforward fashion by
 repeating the calculation above   and  noting that in the general case, 
\seventeenb\   and \dilseven\  are replaced by
\eqn\sanone{{1 \over G+ \Phi} = - \theta + {1 \over g+B},}
and
\eqn\santwo{{\sqrt{det(g+B)} \over g_s} = {\sqrt{det(G+\Phi)} \over G_s}.}

\newsec{Holographic duals}

Gauge invariant operators also appear in the context of
holography. Extending the well known AdS/CFT correspondence
\ref\adscft{J. Maldacena, Adv. Math. Theo. Phys. 2 (1998) 231,
hep-th/9711200; S.S. Gubser, I.R. Klebanov and A.M. Polyakov,
Phys. Lett. 428B (1998) 105, hep-th/9802109; E. Witten, Adv. Theo.
Math. Phys. 2 (1998) 253, hep-th/9802150.}, it has been proposed in
\ref\hi{A. Hashimoto and N. Itzhaki, Phys. Lett. B465 (1999) 142,
hep-th/9907166}\ref\mr{J. Maldacena and J. Russo, JHEP 09 (1999) 025,
hep-th/9908134.}that noncommutative gauge theories are holographic
descriptions of string theories living in appropriate backgrounds
. Then the supergravity (or more generally string theory) modes should
be dual to momentum space operators, as emphasized in
\mr,\ref\dkt{S.R. Das, S. Kalyana Rama and S.P. Trivedi, JHEP 03
(2000) 004, hep-th/9911137.},\dg. Naturally these momentum space
operators should be related to the set of gauge invariant operators
discussed above \drey,\ghi,\dw.  In fact for $d = 3$ it has been
argued in \drey\ that the relationship between the momentum and the
extent in the noncommutative directions encoded in the defintion of
the open Wilson loop operators is visible in dual supergravity.
Similarly \ghi\ argue that the universal large momentum behavior of
the operators \four\ is in agreement with similar behavior in dual
supergravity found in \mr. In fact supergravity predicts an interesting
crossover in the behavior of {\it closed} Wilson loops
\ref\dkita{A. Dhar and Y. Kitazawa, hep-th/0010256.}

In principle these operators are logically distinct from operators
coupled to linearized supergravity about flat space, though in many cases, they are
related to the operators obtained by lineraization around the {\it
background geometry}\ref\relation{See e.g.  S.R. Das and S.P. Trivedi,
Phys.Lett. B445 (1998) 142, hep-th/9804149; S. Ferrara, M. A. Lledó,
A. Zaffaroni, Phys.Rev. D58 (1998) 105029, hep-th/9805082.;
I.Y. Park, A. Sadrzadeh and T.A. Tuan,
hep-th/0010116.} Moreover
as argued in \watimark,
it is possible to obtain the correlation function of the
holographically dual operators from those of the operators obtained by
coupling to linearized supergravity (around {\it flat} spacetime) by
solving the scattering problem in the full geometry.

Consider a non-commutative Yang Mills theory in $p+2n+1$ dimensions. 
We remind the reader that in our notation, (see comments following
equation \one) the non-commutativity 
parameter has rank $2n$; $p$ denotes the remaining spatial directions with 
no noncommutativity. In the following we will define
\eqn\twosix{d = p+2n}
to save clutter in the formulae. 

\subsec{Dual backgrounds}

The supergravity duals were discussed in \hi,\mr, 
\ref\AO{M. Alishalia, Y. Oz and M. M. Sheikh-Jabbari, hep-th/9909215},
\roylu.
The metric in these backgrounds are given 
by:
\eqn\bac{
ds^2=({r\over R})^{({7-d \over 2})}\bigl (-dt^2 +\sum_{i=1}^{p}dx_i^2+
\sum_{i \ \  {\rm odd} }^{2n-1} h_i(dy_i^2+dy_{i+1}^2) \bigr )
+({R \over r})^{({7-d \over 2})}(dr^2 + r^2 d\Omega_{8-p-2n}^2). }
where
\eqn\defr{r^2=\sum_{i=2n+1}^{9-p}y_iy^i}
is the radial coordinate in the $9-p-2n$ directions transverse to the 
brane and 
\eqn\defb{\eqalign{h_i=&{1 \over 1+ a_i^{7-d} r^{7-d}} \cr
a_i^{7-d}=&{b_i^2 \over R^{7-d} l_s^4} \cr
R^{7-d}=&(4\pi)^{({7-d-2 \over 2})}\Gamma({7-d\over 2}) \  g_s N 
l_s^{(7-d)} \prod_{i {\rm  \ \ odd}}
b_i. }}
$N$ above refers to the number of $p+2n$ branes.
Similarly the two-form NS field $B$ and the dilaton are 
\eqn\bdil{\eqalign{
B_{i,i+1}=&{l_s^2 \over b_i} {a^{7-d}r^{7-d} \over 1+ a^{7-d} 
r^{7-d}}, \ \  i=\{1, 3, \cdots, 2n-1 \} \cr
e^{2\phi}=&g_s^2 ({R^{7-d} \over r^{7-d}})^{({7-d-4 \over 2})}
\prod_{i \ \ {\rm odd}}{b_i \over l_s^2} h_i }} 

Note that in \bac\ $x_i, i=1, \cdots p$ and $y_i=1, \cdots 2n$ denote the 
$p+2n$ directions parallel to the brane. The corresponding gauge theory  
also lives in $p+2n$ space directions with non-commutativity parameters 
turned on along  the $2n$ directions, $y=1, \cdots 2n$. 
Let us now consider what happens to the metric in the asymptotic
region, $a_i r \gg 1, i=\{1, 3 \cdots 2n-1\}$. In this region
\eqn\assb{h_i \rightarrow {1 \over a^{7-d} r^{7-d}}=
{R^{7-d} \over r^{7-d }}{l_s^4 \over b_i^2}.}
Rescaling $y_{i,i+1} \rightarrow {b_{i} \over l_s^2} y_{i,i+1}$ then
gives  \bac\ 
\eqn\metricb{ds^2=({r\over R})^{7-d \over 2}\bigl 
(-dt^2 +\sum_{i=1}^{p}dx_i^2 \bigr ) + ({R \over r})^{7-d \over 2}
\bigl(\sum_{i=1}^{2n} dy_i^2 + dr^2 + r^2 d\Omega_{8-p-2n}^2 \bigr).}
The NS field, $B$,  goes to constant asymptotically and the dilaton
is given by
\eqn\assdil{e^{2\phi}=g_s^2 ({ R^{7-p} \over r^{7-p} })^{{3-p \over 2}}.}

In comparison the metric and dilaton background dual to a $p$ dimensional 
ordinary gauge theory are 
\eqn\ordbac{\eqalign{ds^2=& H^{-1/2} (dt^2+\sum_{i=1}^p dx_i^2) + 
H^{1/2}(dy_i^2) \cr
e^{2\phi}&=g_s^2 H^{{3-p\over 2}}.}}
Here, $H$ denotes the appropriate harmonic function which in general
can depend on the $9-p$ transverse coordinates, $y_i$. When the $p$
branes are uniformly distributed in $2n$ of these $9-p$ transverse
directions the harmonic function is given by
\eqn\multi{\eqalign{H & ={R_p^{7-p} \rho \over r^{7-p}} \cr
{\rm where} \ \ R_p^{7-p} & = (4\pi)^{{5-p \over 2}}\Gamma({7-p \over 2})
g_s l_s^{7-p}. }}
In   \multi\  $\rho$  is the number density of $p$ branes  along the
$2n$ directions and $r$ is the transverse distance in the remaining $9-p-2n$
transverse directions.

Comparing \metricb\ \assdil\ with \ordbac\ \multi\ shows that they are
identical, with, 
\eqn\density{\rho={1 \over (4\pi)^{n}} { \Gamma({m
\over 2}) \over \Gamma({7-p \over 2})} { N \over l_s^{4n}} \prod_{i \
\ {\rm odd}} b_i.}

In short, asymptotically, the background \bac\ \bdil, becames
identical to multiceneterd version of the dual for an ordinary $p+1$
dimensional Yang Mills theory, with the branes distributed uniformly
along $2n$ transverse directions.  This behavior is in accord with our
description in section 2 of the non-commutative $p+2n+1$ diemsnional
gauge theory as a particular {\it state} in the $p$ dimensional
ordinary Yang Mills theory. In the $N \rightarrow \infty$ limit,
matrices which satisfy \seven\ have eigenvalues which are uniformly
distributed between $-\infty$ and $+\infty$. This is in agreement with
what we have found above where the branes are uniformly distributed in
the $2n$ directions.

The discussion above implies that supergravity modes which are
perturbations about \bac\ \bdil\ must asymptotically map in a one to
one manner to modes about the background \metricb\ \assdil. The latter
background is considerably simpler, and being the dual of the
multicentered ordinary Yang Mills theory, in some cases better
understood. This simplies the task of classifying the sugra modes in
the noncommutative background.

\subsec{Normal modes and dual operators}

The background \bac\-\bdil\ has nonzero values for several of the
supergravity fields. As a result the analysis of small fluctuations
around such a background is rather complicated and it is difficult
to find normal modes which satisfy decoupled equations. On the other
hand these normal modes should be dual to independent gauge invariant
operators of the holographic theory on the boundary. 

One  such mode is known for supergravity duals of
$3+1$ dimensional NCYM with noncommutativity matrix of rank 2. In the
notation of \bac\ we now have $p=1,n=1$ and the noncommutativity is in
the $(y_1,y_2)$ direction. In that case denote the component of the
ten dimensional graviton polarization along the $(t,x_1)$ directions
by $h_{tx_1}(k_\mu;k_i;k_a)$. We have used the
notation of section 3 : $k_\mu$ denotes momenta along the commutative
directions $t,x_1$, $k_i, i=1,2$ denote momenta along the noncommutative
directions $y_1,y_2$ and $k_a, a= 1, \cdots 6$ denote the  momenta transeverse to the 
three-brane  which may
be also written in terms of a radial momentum along $r$ and the
angular momenta on the $S^5$. When $k_a = 0$  the angular momentum
along the $S^5$ is zero and this is a decoupled mode \mr.  To extract the
dual operator in the gauge theory, we use the fact that the asymptotic
geometry is that of an infinite set of $D1$ branes along $x_1$ which
are smeared in the two transverse directions $y_1,y_2$. In the $D1$
brane the operator which couples to the s-wave graviton with
polarization along $(t,x_1)$ would be given by 
\eqn\threetwo{ \int
d^2\xi ~{\rm Tr}~[e^{ik_i\bX^i}~{\bf \cT_{tx_1}}]} 
where $\cT_{tx_1}$ is the operator whose trace gives the energy
momentum tensor component $T_{tx_1}$. The exponential factor gives the
the operator a $R$-charge or equivalently momentum along the
directions $y_1,y_2$.

Such decoupled modes are, however, rare. In general it is rather
difficult to find these from the supergravity equations. The
observation of the previous subsection, however, relates this
problem to a possibly easier problem of decoupling the equations
around the background of a set of lower dimensional branes with
no B fields. It would be intersting to see whether this approach is
indeed fruitful.

\newsec{Conclusions}

We have proposed a definitive way to identify operators which couple
supergravity modes to noncommutative branes. These are operators
smeared along {\it straight} Wilson tails. It is gratifying that these
operators involve the simplest form of nonlocality required by gauge
invariance.  We have tested our proposal in a rather simple setting,
viz. for an abelian theory and in the DBI approximation. Contrary to
naive expectations even this test is rather nontrivial.  We expect the
couplings of supergravity modes to non-commutative branes found in
this paper to be true in general, beyond these approximations as well,
since it only relies on the construction of non-commutative branes
theory from lower dimensional non-abelian branes.  It is important to
check this by comparing with direct string amplitude computations.

By the nature of our construction, we obtain the operators in the
$\Phi = -B$ description of the noncommutative gauge theory. 
The operators in some other description may be in principle obtained
by using the Seiberg Witten map between these two descriptions.
In fact, in the DBI approximation we have argued
that the operators in any other description may be written down
by a straightforward replacement of the parameters. It is not clear
what happens beyond the DBI approximation. One possibility is to
consider the Seiberg-Witten low energy limit. In this case, one may
hope to obtain the operators in some other description - in particular
the $\Phi = 0$ description by using the Seiberg Witten map in a 
low energy approximation. It would be interesting to see whether the
resulting operators have again a simple form.

One might hope that our proposal can be used to identify the operators
which are involved in the holographic map as well. Here the fact that
the supergravity duals asymptote to geometries which are those of
lower dimensional branes smeared over some of the directions may be
helpful.

\newsec{Appendix}

In this appendix we give the details of the calculations which lead
to the result \dilfifteen\ both from the expansion of \dilsix\ and
\dilseventeen.

\noindent We will use the following properties

\item{(1)} The noncommutative field strength may be written in terms
of the $\sp$ product as in \dilfourteen.

\item{(2)} The $\sp$ product is commutative

\noindent Furthermore since we are dealing with the DBI approximation 
and working
to only $O(A^2)$

\item{(3)} In terms which are $O(A)$ we have to keep the full
expression for $F_{ij}$ as in \dilfourteen. However in terms which
are $O(A^2)$ we can replace $F_{ij}$ by $\partial_i A_j-\partial_j A_i$.

\item{(4)} In terms which involve only the $F_{ij}$ with no explicit
$A_i$ we can replace the $\star$ and $\sp$ product by ordinary products.

\noindent In the following we will refer to these as rules (1)-(4) 
respectively.

Consider first the expansion of the integrand of \dilsix, which we
denote by $I_{com}$
\eqn\apone{I_{com} = {{\sqrt{{\rm det}
(g + B)}}\over g_s}[1 + {{1\over 2}} N^{ij}f_{ji} -
{1\over 4} N^{ij}f_{jk}N^{kl}f_{li}+ {1\over 8} N^{ij}f_{ji}N^{kl}f_{lk}]
+ O(f^3)}
where we have defined
\eqn\aptwo{N^{ij} = ({1 \over g+B})^{ij}}
We have to now use \dilseven\ to write
\eqn\apthree{ N^{ij} = M^{ij} + \theta^{ij}}
where $M$ is defined in \dilsixteen.
Now use the Seiberg-Witten map to express this in terms of $F_{ij}$.
In the second term of \apone\ we have to use equation \plnine\ to get
\eqn\apfour{\eqalign{{{1\over 2}} N^{ij}f_{ji} =
& {{1\over 2}} M^{ij}F_{ji} + {{1\over 2}} \theta^{ij}[(\partial_j A_i - \partial_i A_j)
+\theta^{kl}(\partial_k A_j \sp \partial_l A_i)]\cr
& + {{1\over 2}} M^{ji}\theta^{kl}(A_k \sp \partial_l F_{ij} - F_{ik} \sp F_{jl})\cr
& + {{1\over 2}} \theta^{ji}\theta^{kl}
(A_k \sp \partial_l F_{ij} - F_{ik} \sp F_{jl})+ O(A^3)}}
In the first line we have kept the $F_{ji}$ as it is when it multiplies the
matrix $M$, but have used the expansion in terms of $A_i$ in
\dilfourteen\ when it multiplies $\theta$. The reason will become clear
soon. Using the observations (1)-(4) we can now write this (after some
simplification)
\eqn\apfoura{\eqalign{{{1\over 2}} N^{ij}f_{ji} =
& {{1\over 2}} (M^{ij}F_{ji}) +  \theta^{ij}(\partial_j A_i)\cr
& + {{1\over 2}} M^{ji}\theta^{kl}(A_k \sp \partial_l F_{ij})
+ {{1\over 2}} (M^{ji}F_{ik}\theta^{kl}F_{lj})\cr
& + {{1\over 2}} \theta^{ji}\theta^{kl}
(A_k \sp \partial_l F_{ij}) + {1\over 2}\theta^{ij} \theta^{kl}
(\partial_k A_j \sp 
\partial_l A_i ) -{1 \over 2} \theta^{ji} \theta^{kl} F_{ik} 
F_{jl}
+ O(A^3)}}
In the third and fourth terms of \apone\ already contain two powers of
$F$. Thus to $O(A^2)$ we can set $f_{ij} = F_{ij}$ and we get
\eqn\apfour{\eqalign{-{1\over 4} N^{ij}f_{jk}N^{kl}f_{li}
& = -{1\over 4}(M^{ij}F_{jk}M^{kl}F_{li})-{1\over 4}(\theta^{ij}F_{jk}
\theta^{kl}F_{li}) - {{1\over 2}} (M^{ij}F_{jk}\theta^{kl}F_{li})\cr
{1\over 8} (N^{ij}f_{ji})(N^{kl}f_{lk}) & = {1\over 8}(M^{ij}F_{ji})
(M^{kl}F_{lk})
+ {1\over 8}(\theta^{ij}F_{ji})(\theta^{kl}F_{lk})
+ {1\over 4}(M^{ij}F_{ji})(\theta^{kl}F_{lk})}}
Adding these various contributions and using the realation between $g_s$ and
$G_s$ in \seventeena\ we get \dilfifteen.

Now consider the expansion of the integrand of \dilseventeen\ 
\eqn\dilseventeena{I_{nc}= {1\over G_s}[
P(x) + \theta^{ij}\partial_j(P (x) \sp A_i) + 
{{1\over 2}} \theta^{ij}\theta^{kl}
\partial_j \partial_l [P(x)A_i(x)A_k(x)]_{*3}]}
where
\eqn\dileighteena{P(x) = {\sqrt{{\rm det}(G-B+F(x))}}}
In the first term of \dilseventeena\ we will need the expansion of the
determinant to $O(F^2)$, in the second term we need the expansion to
$O(F)$ and in the last term to $O(F^0)$. 
The first term becomes
\eqn\apfive{P(x) = {{\sqrt{{\rm det}(G-B)}}}[1 + {{1\over 2}} (M^{ij}F_{ji})
-{1\over 4}(M^{ij}F_{jk}M^{kl}F_{li}) + {1\over 8}(M^{ij}F_{ji})
(M^{kl}F_{lk})]}
We have replaced star products by ordinary products in accordance to
our rule (4) above.
The second term in \dilseventeena\ becomes
\eqn\apsix{\eqalign{\theta^{ij}\partial_j(P \sp A_i) =
{{\sqrt{{\rm det}(G-B)}}}&[\theta^{ij}\partial_j A_i + {{1\over 2}} M^{kl}\theta^{ij}
(\partial_j F_{lk} \star^\prime A_i)\cr
&+ {{1\over 2}} M^{kl}\theta^{ij}(F_{lk}\sp \partial_jA_i)]}}
Using rules (2)-(4) above this may be written as
\eqn\apsixa{\eqalign{\theta^{ij}\partial_j(P \sp A_i) =
{{\sqrt{{\rm det}(G-B)}}}&[\theta^{ij}\partial_j A_i + {{1\over 2}} M^{kl}\theta^{ij}
(\partial_j F_{lk} \star^\prime A_i)\cr
&+ {1\over 4} (M^{kl}F_{lk})(\theta^{ij}F_{ji})]}}
Finally we consider the
third term in \dilseventeena. Here we can replace $P$ by
${\sqrt{{\rm det}(G-B)}}$ since the other terms can contribute only
$O(A^3)$ terms. Then the triple product collapses to a $\star^\prime$
product by virtue of \plfivea. Using this one finds
\eqn\apsevena{\eqalign{{{1\over 2}} \theta^{ij}\theta^{kl}
\partial_j \partial_l [PA_iA_k]_{*3} & = 
{{1\over 2}}{\sqrt{{\rm det}(G-B)}}
\theta^{ij}\theta^{kl}[\partial_l\partial_jA_i \sp A_k
+ A_i \sp \partial_l\partial_jA_k \cr
& + \partial_j A_i \sp \partial_l A_k + \partial_l A_i \sp \partial_j A_k]}}
Since this is already $O(F^2)$ we can use rule (3) above and then use
the rule (4) to write
\eqn\apseven{\eqalign{{{1\over 2}} \theta^{ij}\theta^{kl}
\partial_j \partial_l [PA_iA_k]_{*3} & = 
{\sqrt{{\rm det}(G-B)}}[
{{1\over 2}} \theta^{ij}\theta^{kl}
(\partial_l F_{ji} \sp A_k)\cr
& + {1\over 8} (\theta^{ij}F_{ji})(\theta^{kl}F_{lk}) 
+ {1 \over 2} \theta^{ij} \theta^{kl} (\partial_l A_i \sp \partial_j A_k)
+ O(A^3)]}}
Adding the contributions \apfive,\apsix and \apseven\ 
and using the commutative nature of the $\sp$ product 
one again gets \dilfifteen.

\newsec{Acknowledgements} We would like to thank S. Mukhi for discussions.
We thank M. van Raamsdonk for comments about the first version of this
paper, especially for pointing out the importance
of symmetrized traces in supergravity coupling to branes.
\listrefs
\end